\documentclass[twocolumn,aps,prc,superscriptaddress,nofootinbib,showpacs,floatfix,longbibliography]{revtex4-1}
\usepackage{url}
\usepackage{cancel}
\usepackage[colorlinks,linkcolor=blue,citecolor=blue,filecolor=black,urlcolor=blue]{hyperref}
\usepackage{epsfig,graphics}
\usepackage{graphicx}
\usepackage{dcolumn}
\usepackage{bm}
\usepackage[usenames]{color}
\usepackage{amssymb}
\usepackage{amsmath}
\usepackage{multirow}
\usepackage{float}
\usepackage{harpoon}
\usepackage{MnSymbol}
\usepackage{appendix}
\usepackage{color}
\usepackage{hyperref}
\usepackage{cleveref}

\newcommand{\nch} {N_{\mathrm{ch}}}
\newcommand{\nchrec} {N_{\mathrm{ch}}}
\newcommand{\sqrtsnn}{\mbox{$\sqrt{s_{\mathrm{NN}}}$}}
\newcommand{\pT} {p_{\mathrm{T}}}
\newcommand{\lr}[1]{\left\langle #1\right\rangle}

\usepackage{CJK}
\hfuzz=\maxdimen
\begin{document}
\title{Impact of event activity variable on the ratio observables in isobar collisions}
\newcommand{\sbu}{Department of Chemistry, Stony Brook University, Stony Brook, NY 11794, USA}
\newcommand{\bnl}{Physics Department, Brookhaven National Laboratory, Upton, NY 11976, USA}
\newcommand{\ucla}{Department of Physics and Astronomy, University of California, Los Angeles, California 90095, USA}
\author{Jiangyong Jia}\affiliation{\sbu}\affiliation{\bnl}
\author{Gang Wang}\affiliation{\ucla}
\author{Chunjian Zhang}\affiliation{\sbu}

\begin{abstract}
The STAR isobar data of $^{96}$Ru+$^{96}$Ru and $^{96}$Zr+$^{96}$Zr collisions at $\sqrt{s_{\mathrm{NN}}}= 200$ GeV show that ratios of observables ($R_{\mathcal{O}}$) such as the multiplicity distribution, $p(N_{\mathrm{ch}})$, and the harmonic flow, $v_n$, deviate from unity, when presented as a function of centrality, $c$~\cite{STAR:2021mii}. These deviations have been attributed to the differences in the shape and radial profiles between $^{96}$Ru and $^{96}$Zr nuclei. In addition, the ratios $R_{\mathcal{O}}(x)$ depend on the choice of the event activity variable $x$, which could be either $N_{\mathrm{ch}}$  or centrality.  We estimate the difference $\Delta R$ between these two choices, based on the published $p(N_{\mathrm{ch}})$, as well as those from a multiphase transport (AMPT) model with varied nuclear structure parameters: nuclear radius ($R_0$), surface diffuseness ($a_0$), quadrupole deformation ($\beta_2$), and octupole deformation ($\beta_3$). In contrary to $R_{v_n}(c)$, $R_{v_n}(N_{\mathrm{ch}})$ is nearly independent of the analysis approaches, suggesting that nonflow effects are better controlled by $N_{\mathrm{ch}}$ than $c$. The ratios of observables sensitive to the chiral magnetic effect (CME) are also much closer to unity for $x=N_{\mathrm{ch}}$ than $x=c$, indicating that the ratios calculated at the same $N_{\mathrm{ch}}$ provide a better baseline for the non-CME background. According to the AMPT results, the dominant parameter for $\Delta R$ is $a_0$, while $R_0$ and $\beta_n$ are only important in central collisions. The published $p(N_{\mathrm{ch}})$ is also used to estimate $\Delta R_{\left\langle p_{\mathrm{T}}\right\rangle}$ for mean transverse momentum, which is non-negligible compared with $R_{\left\langle p_{\mathrm{T}}\right\rangle}-1$.
\end{abstract}
\maketitle

The main goal of the isobar collisions ($^{96}$Ru+$^{96}$Ru and $^{96}$Zr+$^{96}$Zr  at $\sqrtsnn=200$ GeV) at the Relativistic Heavy Ion Collider (RHIC) is  to search for the ellusive chiral magnetic effect (CME)~\cite{Fukushima:2008xe,STAR:2009wot}. Since $^{96}$Ru contains more protons than $^{96}$Zr, $^{96}$Ru+$^{96}$Ru collisions generate stronger initial magnetic fields than $^{96}$Zr+$^{96}$Zr collisions, leading to potentially detectable differences in the CME sensitive observables, such as the $\Delta \gamma_{112}$ correlator~\cite{Voloshin:2010ut}. However, no evidence of the CME signal has been found in the first isobar analysis from the STAR Collaboration~\cite{STAR:2021mii}. Instead, the search uncovers significant differences in many observables, supportive of the structure differences between the two nuclei~\cite{Jia:2021oyt}. In particular, the  collective-flow results reveal large deformations in the two nuclei~\cite{Zhang:2021kxj}, as well as a difference in the neutron skin~\cite{Xu:2021vpn}. This opens up a new opportunity to probe the collective nuclear structure using high-energy isobar collisions~\cite{Jia:2021tzt}. Ultimately,  the effects of nuclear structures~\cite{Li:2022bhl} and the nonflow backgrounds~\cite{Feng:2021pgf} have to be accounted for, before  any residual differences might be attributable to the genuine CME signal.  

The collective shape and radial profile of atomic nuclei are often described by a deformed Woods-Saxon (WS) density, separately for protons and neutrons,
\begin{align}\label{eq:f1}
\rho(r,\theta,\phi)&\propto\frac{1}{1+e^{[r-R_0\left(1+\beta_2 Y_2^0(\theta,\phi) +\beta_3 Y_3^0(\theta,\phi)\right)]/a_0}}.
\end{align}
The nuclear structure parameterization includes  half-density radius ($R_0$),   surface diffuseness ($a_0$), and  axial symmetric quadrupole deformation ($\beta_2$) and octupole deformation ($\beta_3$). Model studies of elliptic flow ($v_2$) and triangular flow ($v_3$) suggest   $\beta_2\sim0.15$ for $^{96}$Ru and   $\beta_3\sim0.2$ for $^{96}$Zr, respectively~\cite{Zhang:2021kxj}. In the mid-central collisions, the comparison of $v_2$ and the multiplicity distribution between the two isobaric systems supports the idea that $^{96}$Zr has a thicker neutron skin than $^{96}$Ru~\cite{Li:2019kkh,Jia:2021oyt}. Predictions for other observables and their sensitivities to deformation and neutron skin have also been made, such as mean transverse momentum $\lr{\pT}$~\cite{Xu:2021uar}, $\lr{\pT}$ fluctuations~\cite{Jia:2021qyu}, and $v_n$--$\pT$ correlations~\cite{Giacalone:2019pca,Bally:2021qys,Jia:2021wbq}.

\begin{figure}[!h]
\includegraphics[width=0.8\linewidth]{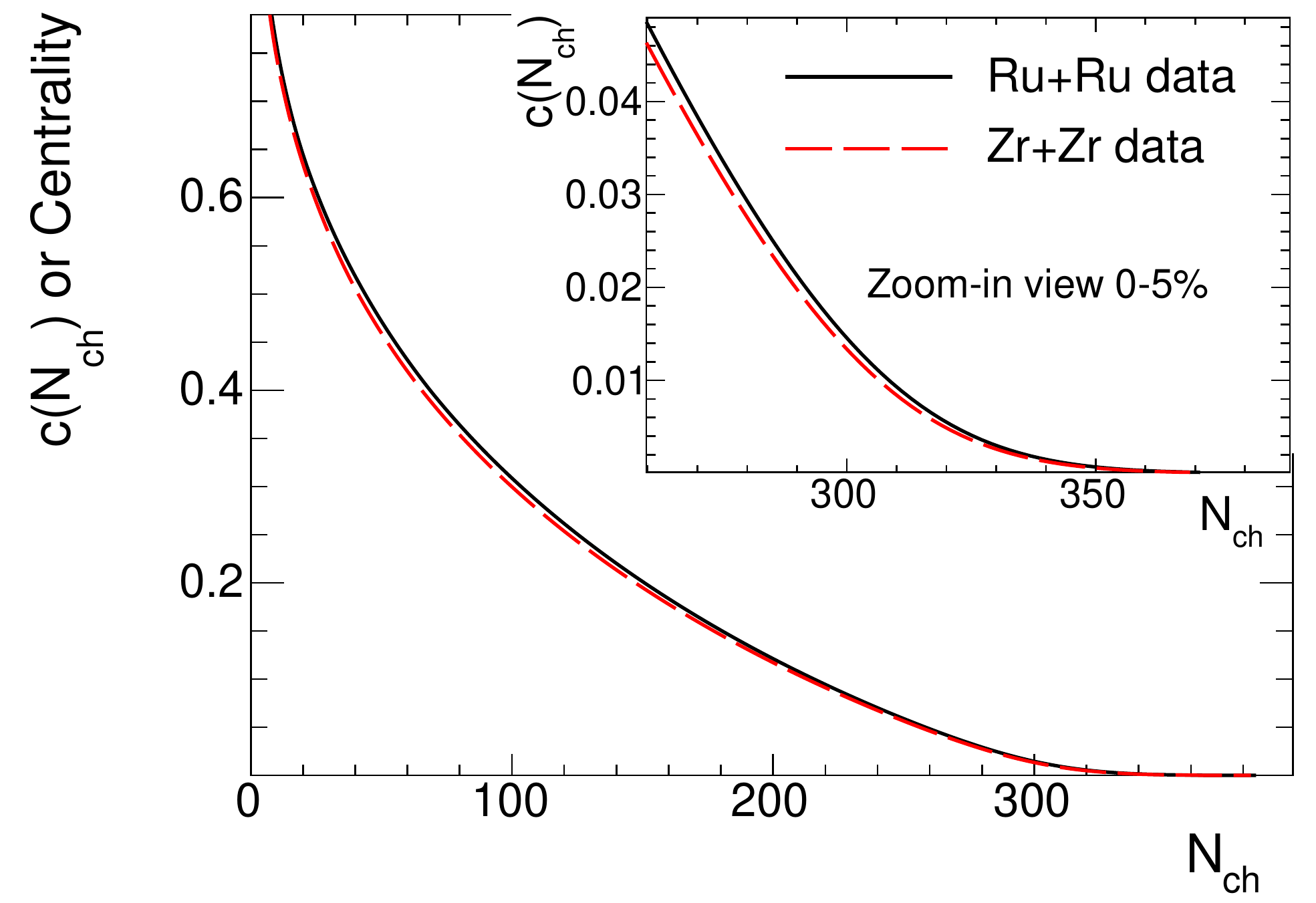}
\vspace*{-.3cm}
\caption{\label{fig:0} Centrality vs $\nchrec$, i.e. $c(\nchrec)$ obtained from the published $p(\nchrec)$ in $^{96}$Ru+$^{96}$Ru and $^{96}$Zr+$^{96}$Zr collisions~\cite{STAR:2021mii}. Similar relations are also obtained in the AMPT model for each variation of WS parameters (not shown).}
\end{figure}

One thing to note is that the observable ratios between the two isobaric systems are usually presented as a function of an event activity variable, which is also affected by the nuclear structure. The STAR measurements use centrality ($c$) and charged-particle multiplicity ($\nchrec$) in the pseudorapidity range of $|\eta|<0.5$  as the event activity variables. Centrality is calculated directly from the $\nchrec$ distribution, $p(\nchrec)$, as $c(\nchrec)=\int_{\nchrec}^{\infty} p(n) d n$. The $c(\nchrec)$ function is monotonic with a range of $0\leq c\leq1$ (see Fig.~\ref{fig:0} for $c(\nchrec)_{\mathrm{Ru}}$ and $c(\nchrec)_{\mathrm{Zr}}$). The STAR data show that $p(\nchrec)$ is broader in Ru+Ru than in Zr+Zr collisions, and thus events in the two collision systems at matching $\nchrec$ correspond to different or mismatched $c$ values, and vice versa. Therefore,  the ratio of a given observable $\mathcal{O}$ relies on the event activity variable $x$: $R_{\mathcal{O}}(x)=\mathcal{O}(x)_{\mathrm{Ru}}/\mathcal{O}(x)_{\mathrm{Zr}}$, with $x=c$ or $\nchrec$. We quantify the effect of mismatched event activity with the first-order approximation, 
\small{
\begin{align}\label{eq:f2a}
&\Delta R_{\mathcal{O}}=\frac{\mathcal{O}\left(c_{\mathrm{Ru}}\right)_{\mathrm{Ru}}-\mathcal{O}\left(c_{\mathrm{Zr}}\right)_{\mathrm{Ru}}}{\mathcal{O}\left(c_{\mathrm{Zr}}\right)_{\mathrm{Zr}}} \approx \frac{\mathcal{O}'(c)}{\mathcal{O}(c)} \Delta c \;,\\\label{eq:f2b}
&\Delta R_{\mathcal{O}} =\frac{\mathcal{O}\left(N_{\mathrm{ch Ru}}\right)_{\mathrm{Ru}}-\mathcal{O}\left(N_{\mathrm{ch Zr}}\right)_{\mathrm{Ru}}}{\mathcal{O}\left(N_{\mathrm{ch Zr}}\right)_{\mathrm{Zr}}} \approx \frac{\mathcal{O}'\left(\nchrec\right)}{\mathcal{O}\left(\nchrec\right)} \Delta \nchrec\;.
\end{align}}\normalsize
Here $\Delta c= c_{\mathrm{Ru}}-c_{\mathrm{Zr}}$, $\mathcal{O}'(c) = d\mathcal{O}(c)/dc$, and similarly for $\Delta \nchrec$ and $\mathcal{O}'\left(\nchrec\right)$. Both $\Delta c$ and $\Delta \nchrec$ can be directly read off from Fig.~\ref{fig:0}. Eq.~\eqref{eq:f2a} provides the correction from the ratio at matching $c$ to the ratio at matching $\nchrec$, while Eq.~\eqref{eq:f2b} provides the correction from the ratio at matching $\nchrec$ to the ratio at matching $c$. Therefore these two corrections have the opposite signs. $\Delta R_{\mathcal{O}}$ is proportional to the derivative $\ln'(\mathcal{O}(c))=\mathcal{O}'(c)/\mathcal{O}(c)$, and hence its sign depends on whether $\mathcal{O}$ increases or decreases with centrality. In the following figures and related discussions, $\Delta R_{\mathcal{O}}$ always denotes the ratio at matching $\nchrec$ minus that at matching centrality, unless specified otherwise.

In this paper, we apply Eqs.~\eqref{eq:f2a} and \eqref{eq:f2b} directly to the published experimental data~\cite{STAR:2021mii} to estimate the change in the ratio when switching the  $x$-axis between $c$ and $\nchrec$. We then investigate the origin of this change using a multiphase transport (AMPT) model~\cite{Lin:2004en}. A previous study~\cite{Jia:2021oyt} explains how $p(\nchrec)$ is influenced by each of the four WS parameters ($R_0$, $a_0$, $\beta_2$, and $\beta_3$). The values of these parameters for the isobaric nuclei are taken from Ref.~\cite{Jia:2021oyt} and listed in Table~\ref{tab:1}. Each scenario of the AMPT study
has a relation between $c$ and $\nchrec$, similar to the real data in Fig.~\ref{fig:0}. These relations in turn can be used as input for Eqs.~\eqref{eq:f2a} and \eqref{eq:f2b}  to estimate the impact of each parameter on $\Delta R_{\mathcal{O}}$, when switching between the centrality dependence and the $\nchrec$ dependence. 

\begin{table}[!h]
\centering
\begin{tabular}{|c|cccc|}\hline 
\text {Species} &\; $R_0$ (fm)\; & \;$a_{0}$ (fm)\;  & $\beta_{2}$ & $\beta_{3}$  \\\hline 
$^{96}$Ru & 5.09  & 0.46   & 0.162 & 0  \\\hline 
$^{96}$Zr & 5.02  & 0.52   & 0.06 & 0.20  \\\hline
\end{tabular}
\caption{\label{tab:1} Collective nuclear structure parameters for $^{96}$Ru and $^{96}$Zr from Ref.~\cite{Jia:2021oyt}.} 
\end{table}

\begin{figure}[!h]
\includegraphics[width=1\linewidth]{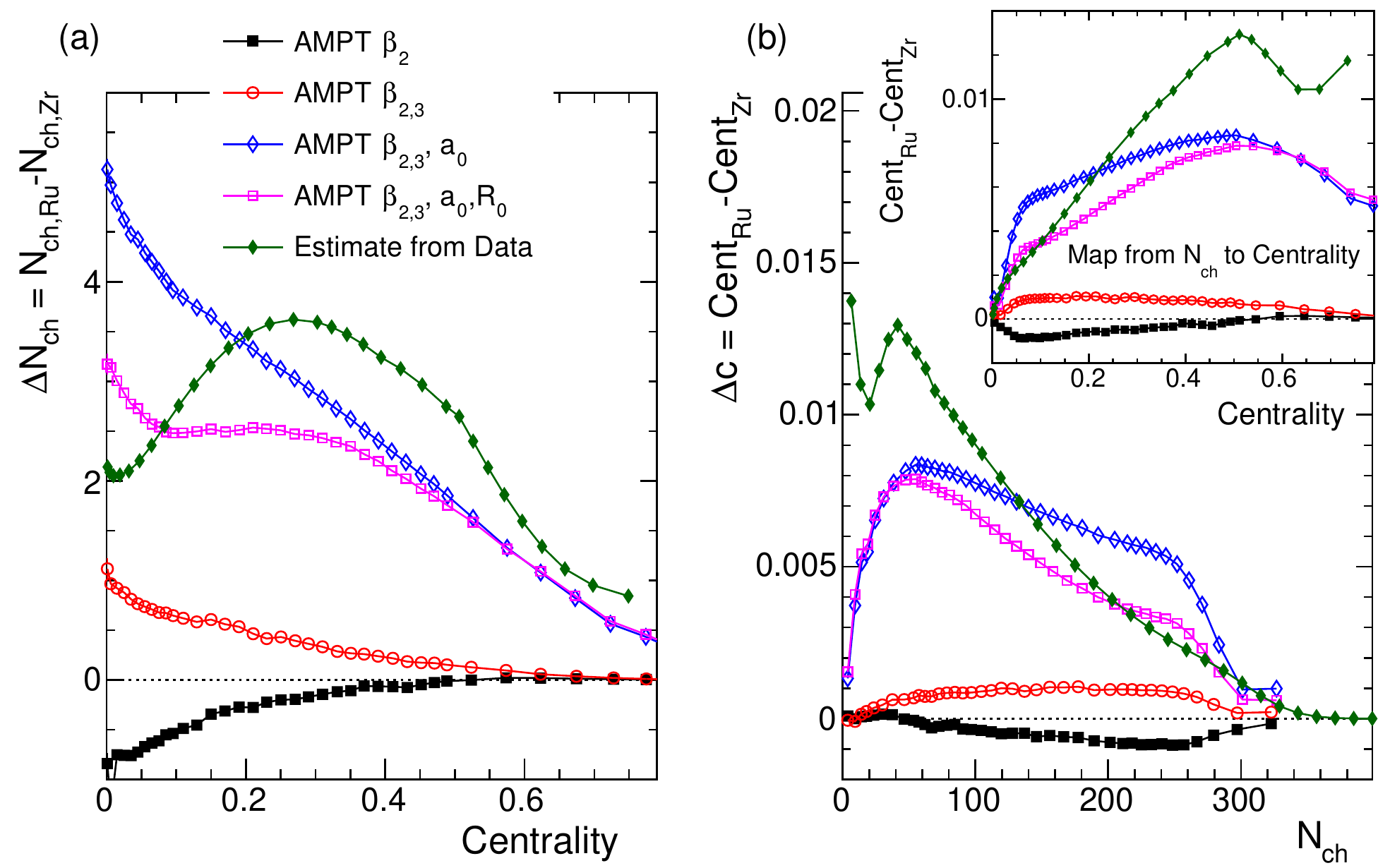}
\vspace*{-.3cm}
\caption{\label{fig:1} (a) $\Delta \nchrec = N_{\mathrm{ch,Ru}}(c)-N_{\mathrm{ch,Zr}}(c)$ for events at matching centrality for isobar collisions. (b) $\Delta c= c_{\mathrm{Ru}}(\nchrec)-c_{\mathrm{Zr}}(\nchrec)$ for events at matching  $\nchrec$. Both are obtained directly from Fig.~\ref{fig:0}. The insert panel shows $\Delta c$ after mapping $\nchrec$ to centrality in the $x$-axis. In comparison with the data, the AMPT calculations are added sequentially to include the effects of $\beta_2$, $\beta_3$, $a_0$ and $R_0$ in the two systems in Tabel~\ref{tab:1}.}
\end{figure}

Figure~\ref{fig:1}(a) shows $\Delta \nchrec (c) = N_{\mathrm{ch,Ru}}(c)-N_{\mathrm{ch,Zr}}(c)$ and Fig.~\ref{fig:1}(b) shows $\Delta c (\nchrec)=c_{\mathrm{Ru}}(\nch)-c_{\mathrm{Zr}}(\nch)$ as a function of centrality. Note that $\Delta \nchrec$ is calculated at matching $c$, and $\Delta c$ is calculated at matching $\nchrec$, whose values can be read off directly from Fig.~\ref{fig:0} or from analogue plots in the AMPT simulations~\cite{Jia:2021oyt}. Furthermore, the $x$-axis in all following figures, either centrality or $\nch$, are always chosen to represent the values obtained for Ru+Ru collisions. Both $\Delta \nchrec$ and $\Delta c$ can then be presented as a function of either $c$ or $\nchrec$. Since the $p(\nchrec)_{\mathrm{Ru}}$ distribution is broader than $p(\nchrec)_{\mathrm{Zr}}$, events with the same $\nchrec$ correspond to larger $c$ values (or more peripheral collisions) in Ru+Ru than in Zr+Zr. Conversely, events with the same $c$ correspond to larger $\nchrec$ in Ru+Ru than in Zr+Zr. The overall centrality dependence of $\Delta \nchrec$ and $ \Delta c$ in Fig.~\ref{fig:1} can be qualitatively reproduced by the AMPT model after taking into account the influences of all four WS parameters. A previous study~\cite{Jia:2021oyt} has shown that the impact of the four WS parameters on isobar ratios are independent of each other when chosen from Table~\ref{tab:1}. Therefore, the impact of $\beta_2$ is reflected by the points represented by label ``$\beta_{2}$'',  the impact of $\beta_3$ is reflected by the change from label ``$\beta_{2}$'' to label ``$\beta_{2,3}$'', the impact of $a_0$ is reflected by the change from label ``$\beta_{2,3}$'' to label ``$\beta_{2,3},a_0$'', and so on. We briefly summarize the AMPT results as follows: 1) The diffuseness parameter $a_0$ has the largest impact on $p(\nchrec)$, and the smaller $a_0$ value of Ru tends to make  $p(\nchrec)_{\mathrm{Ru}}$ broader than $p(\nchrec)_{\mathrm{Zr}}$. 2) Half-density radius $R_0$ also plays a sizeable role in central collisions, and the larger $R_0$ of Ru tends to reduce the range of $p(\nchrec)_{\mathrm{Ru}}$ compared with $p(\nchrec)_{\mathrm{Zr}}$. 3) The effects of nuclear deformation are mostly concentrated in central collisions, with $\beta_{2\mathrm{Ru}}$ and $\beta_{3\mathrm{Zr}}$ working in the opposite directions.

Besides $\Delta c$, we also need to know the local slope of $\ln(\mathcal{O}(c))$ in Eq.~\eqref{eq:f2a},  
to estimate $\Delta R_{\mathcal{O}}$ when switching from the ratio at matching $c$ to the ratio at matching $\nchrec$. Ref.~\cite{STAR:2021mii} has published results for $v_2$, $v_3$, and the CME sensitive observables such as $\Delta\gamma_{112}$, $\Delta \delta$, and $\kappa_{112}\equiv\Delta\gamma_{112}/(v_2\Delta\delta)$. We shall not explain the definitions and the physical meanings of these CME related observables (see Ref.~\cite{Li:2020dwr}), but instead just provide an estimate of $\Delta R_{\mathcal{O}}$ and explore its origin with the AMPT model. 

\begin{figure}[!h]
\includegraphics[width=1\linewidth]{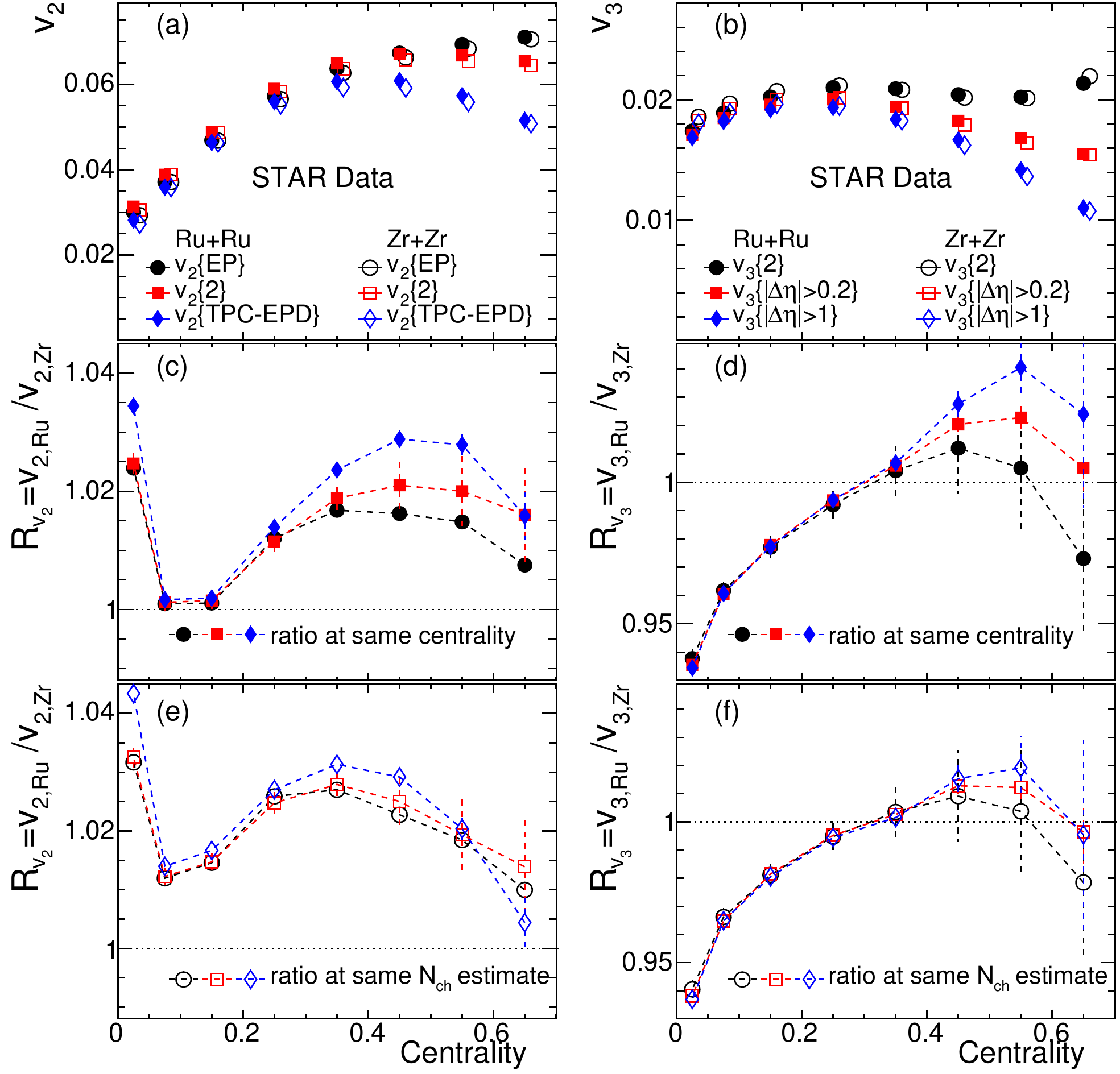}
\vspace*{-.3cm}
\caption{\label{fig:2} The STAR $v_2$ (a) and $v_3$ (b) data from isobar collisions using three analysis methods from Ref.~\cite{STAR:2021mii}. The ratios $R_{v_2}(c)$ (c) and $R_{v_3}(c)$ (d) are calculated at matching centrality directly with the $v_n$ data from the top panels. The ratios $R_{v_2}(\nchrec)$ (e) and  $R_{v_3}(\nchrec)$ (f) are calculated at matching $\nchrec$ using Eq.~\eqref{eq:f2a} with data from Fig.~\ref{fig:1} as described in the text.}
\end{figure}

The $v_2$ and $v_3$ data in Ref.~\cite{STAR:2021mii} involve several analysis methods with different sensitivities to nonflow effects. We choose the STAR results from three methods, and display them in the top panels of Fig.~\ref{fig:2}.
The directly-calculated $R_{v_n}(c)$ ratios are presented as a function of centrality in the middle panels.
The bottom panels show the $R_{v_n}(\nchrec)$ ratios taken at $N_{\mathrm{ch,Ru}}$ for each centrality bin using Eq.~\eqref{eq:f2a} with data from Fig.~\ref{fig:1}. In general, $R_{v_2}(\nchrec)$ is larger than $R_{v_2}(c)$, since $v_2$ increases with $c$, yielding a positive $\ln'(v_2(c))$. In contrast, $R_{v_3}(\nchrec)$ is nearly the same as $R_{v_3}(c)$, as expected from the weak centrality dependence of $v_3$. In peripheral collisions, where $v_2$ and $v_3$ are highly affected by nonflow effects, the ratios are method-dependent as expected. Remarkably, the  $R_{v_n}(\nchrec)$ ratios from different methods are much closer to each other than the case of $R_{v_n}(c)$. This suggests that nonflow effects in the two isobaric systems are controlled by $\nchrec$ instead of centrality. In other words, nonflow contributions in Ru+Ru and Zr+Zr collisions are almost the same at matching $\nchrec$, and hence are different at matching centrality.~\footnote{In this scenario, the remaining differences in $R_{v_n}(\nchrec)$ between the analysis methods could be attributed to the dilution effects associated with the different amounts of nonflow effects in  those methods. However, the STAR measurements need to be repeated with much finer centrality bins, such that Eq.~\eqref{eq:f2a} can provide a more accurate estimate.}

\begin{figure}[!h]
\includegraphics[width=0.95\linewidth]{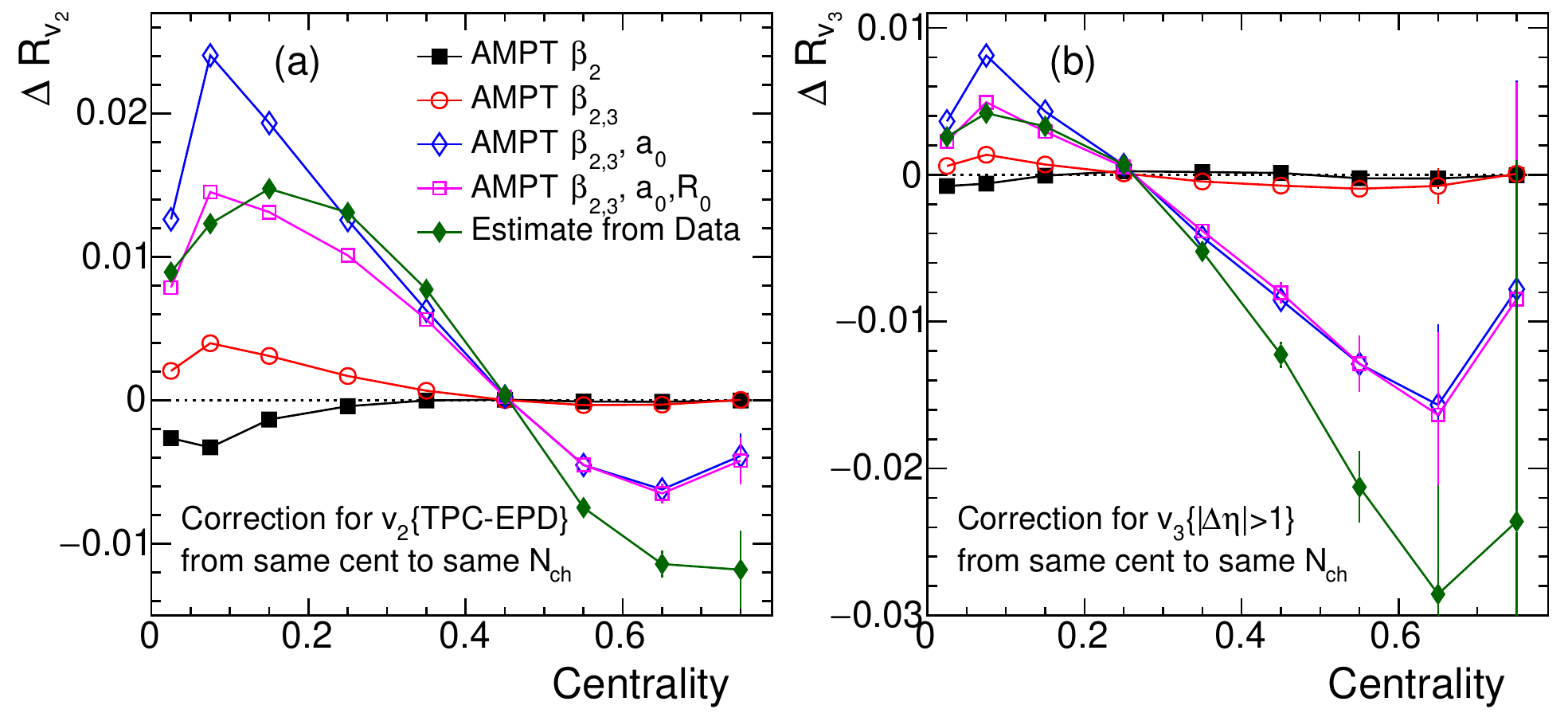}
\vspace*{-.3cm}
\caption{\label{fig:3} The difference between the ratio at matching $\nchrec$ and the ratio at matching centrality:  $\Delta R_{v_2}$ (a) and $\Delta R_{v_3}$ (b),  calculated using $p(\nchrec)$ from the STAR data~\cite{STAR:2021mii} as well as AMPT including the effects of WS parameters in Eq.~\eqref{eq:f1}.}
\end{figure}

Next, we estimate the impact of the WS parameters on the isobar ratio in the AMPT model using Fig.~\ref{fig:1} as input. We select the results whose methods have smaller nonflow, i.e. $v_2\{\mathrm{TPC-EPD}\}$ and $v_3\{|\Delta\eta|>1\}$, and present them in Fig.~\ref{fig:3}. The model calculations are compared with the $\Delta R_{v_n}$ values estimated from the STAR data, which are just the differences between the open diamonds in the bottom panels and the solid diamonds in the middle panels of Fig.~\ref{fig:2}.  As expected, $a_0$ plays a leading role in the difference, followed by $R_0$, and then $\beta_3$ and $\beta_2$. The calculated $\Delta R_{v_n}$ after considering all the nuclear structure effects (open boxes) are similar to the data (solid diamonds) for $c<0.2$, but are smaller in magnitude elsewhere. 

\begin{figure}[!h]
\includegraphics[width=1\linewidth]{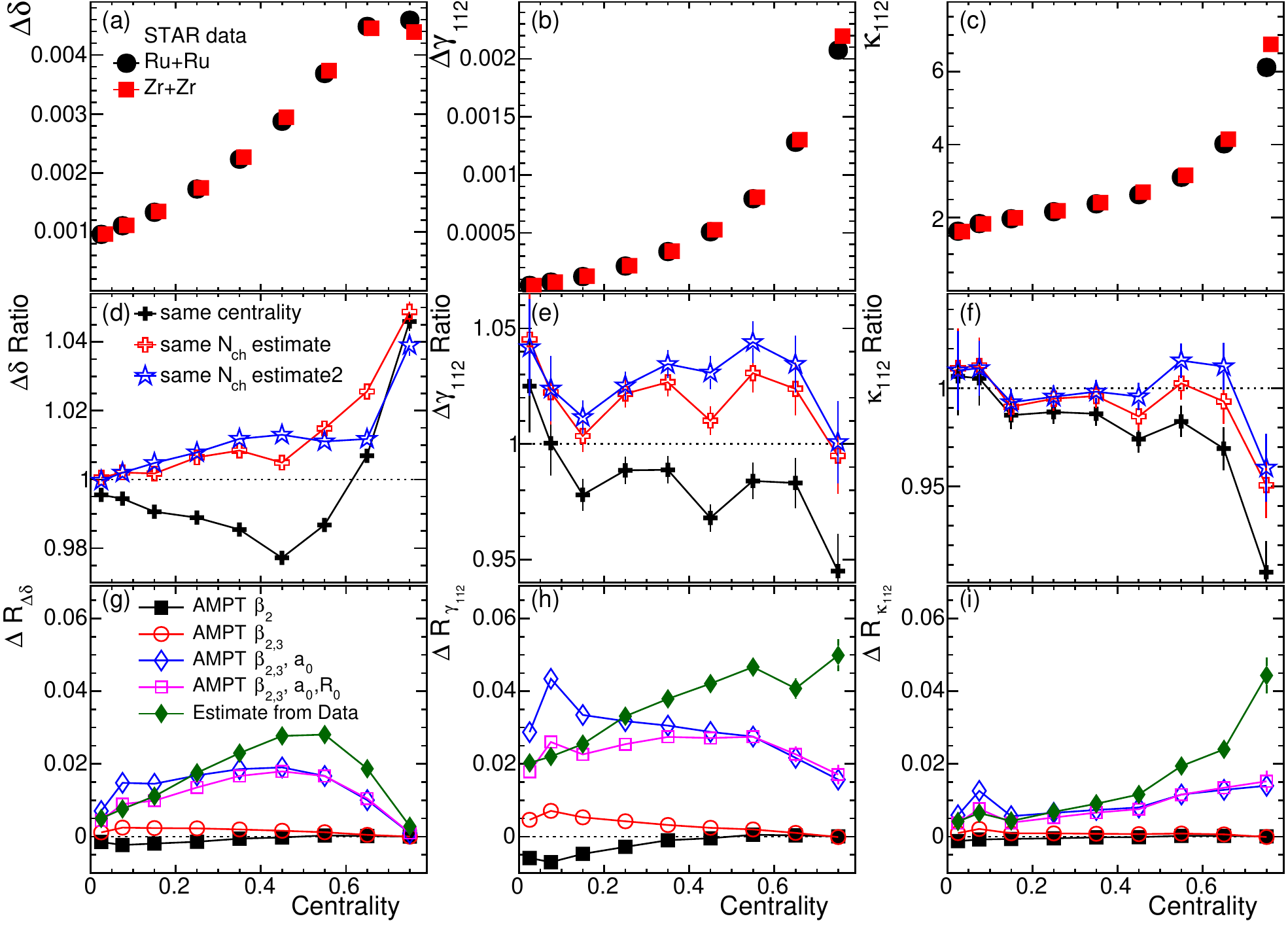}
\vspace*{-.3cm}
\caption{\label{fig:4} The results of the CME sensitive observables $\Delta \delta$ (left column), $\Delta \gamma_{112}$ (middle column) and $\kappa_{112}$ (right column) from isobaric systems. The top row shows the published data. The middle row displays the isobar ratios calculated at matching centrality (filled crosses), and at matching $\nchrec$ estimated via a linear interpolation from Eq.~\eqref{eq:f2a} (open crosses) or via a second-order polynomial interpolation (open stars) labelled as "estimate2". The bottom row shows the $\Delta R$ due to switching the event activity variable, based on $p(\nchrec)$ from the STAR data~\cite{STAR:2021mii} as well as AMPT including the effects of WS parameters in Eq.~\eqref{eq:f1}.}
\end{figure}

Figure~\ref{fig:4} shows the STAR data of the CME sensitive observables: $\Delta\delta$ (a), $\Delta\gamma_{112}$ (b) and $\kappa_{112}\equiv\Delta\gamma_{112}/(v_2\Delta\delta)$ (c). The middle row presents the ratios of Ru+Ru to Zr+Zr for the corresponding observables at matching $c$ and at matching $\nchrec$. In the latter case, we have also tried an alternative approach by modifying Eq.~\eqref{eq:f2a} to a second-order polynomial interpolation, determined by every three adjacent points (labelled as "estimate2"). The results from the two interpolation methods are nearly identical in central collisions, but show some deviations elsewhere. To improve the accuracy of our estimate, STAR measurements need to be repeated with much finer centrality bins in the future. For most centrality intervals under study, both the $\Delta\delta$ ratios and the $\Delta\gamma_{112}$ ratios change from below unity to above unity after switching from matching $c$ to matching $\nchrec$. Such qualitative changes reveal the importance of choosing the proper event activity variable, as this choice may strongly affect the perception of whether/how $R$ deviates from unity, and whether it is attributable to the genuine CME effects. In general, the $\kappa_{112}$ ratios at matching $\nchrec$ are consistent with unity, whereas those at matching $c$ are significantly below unity, suggesting that the non-CME backgrounds in the two systems are better controlled by $\nchrec$ than  $c$. The bottom panels show that the effects of switching the event activity variable tend to be larger in more peripheral collisions, and are dominated by $a_0$.

\begin{figure}[!h]
\includegraphics[width=1\linewidth]{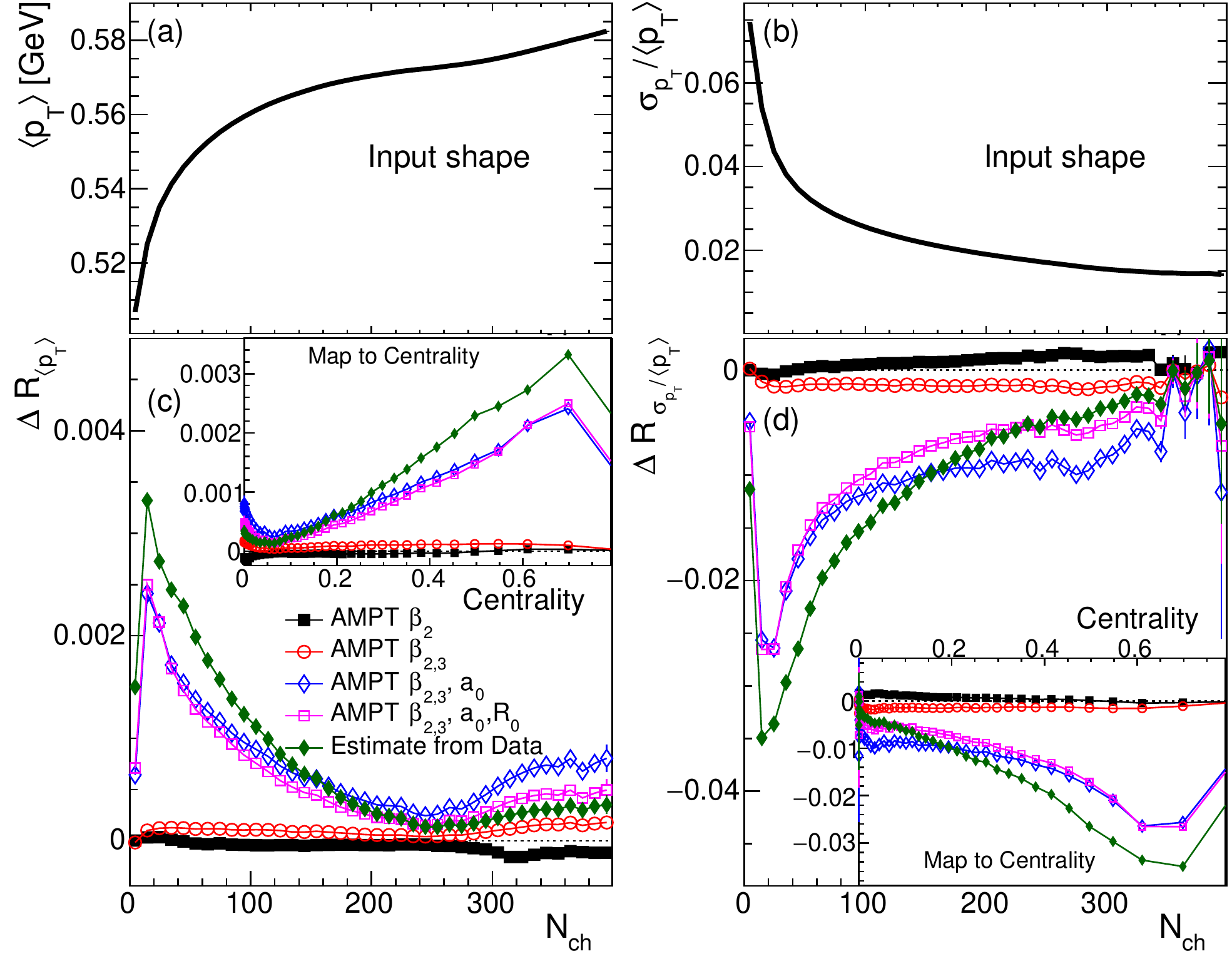}
\vspace*{-.3cm}
\caption{\label{fig:5} The input $\nchrec$ dependence of $\lr{\pT}$ (a) and  $\sigma_{\pT}/\lr{\pT}$ (b), and the corresponding $\Delta R$ (change from the ratio at matching $\nchrec$ to ratio  at matching centrality) based on the $p(\nchrec)$ distributions from STAR data and AMPT. The insert panels show the same $\Delta R$  after mapping $\nchrec$ to centrality in the $x$-axis.}
\end{figure}

The last set of observables to be studied in this paper are $\langle\pT\rangle$ and its fluctuations in terms of scaled-variance, $\sigma_{\pT}/\lr{\pT}$. To estimate $\Delta R_{\mathcal{O}}$ for these observables,  we need to express them as a function of $\nchrec$ (Eq.\eqref{eq:f2b}). Here we assume that the multiplicity dependence of either observable in isobar collisions is similar to that in Au+Au collisions at 200 GeV~\cite{STAR:2019dow}. In particular, scaled-variance is set to follow a power-law dependence, $\sigma_{\pT}/\lr{\pT}\propto (\nchrec)^{-n}$, with $n\sim 0.4$~\cite{ALICE:2014gvd}. Motivated by a hydrodynamics argument~\cite{Gardim:2019brr}, $\lr{\pT}$ undergoes a sharp rise at low $\nchrec$~\cite{ALICE:2013rdo,ALICE:2018hza}, followed by a relatively flat behavior in the mid-central region, and then by a second rise in the ultra-central region with an increase of about 10 MeV. The input distributions are shown in the top row of Fig.~\ref{fig:5}, and are used to estimate the ratio change when switching from matching $\nchrec$ to matching $c$.

The bottom panels in Fig.~\ref{fig:5} show the change of ratios from matching $\nchrec$ to matching $c$ via Eq.~\eqref{eq:f2b}. The insert panel in Fig.~\ref{fig:5}(c) shows the same $\Delta R_{\lr{\pT}}$ but plotted as a function centrality. Since $\lr{\pT}$ increases with $\nchrec$ (or decreases with centrality), the difference $\Delta R_{\lr{\pT}}$ is positive according to Eq.~\ref{eq:f2b}. Note that if we calculate instead the change of ratio from the same $c$ to the same $\nchrec$, as is done in all other figures via Eq.~\eqref{eq:f2a}, the results would have the opposite signs. $\Delta R_{\lr{\pT}}$ reaches a value of $\sim0.0005$ at $c= 0.2$, $\sim0.002$ at $c=0.5$, and even larger in more peripheral collisions. This difference is potentially a significant fraction of the signal, $R_{\lr{\pT}}-1$, as a function of centrality, which is predicted to be $\sim0.003$ at $c= 0.2$ and $\sim0.005$ at $c=0.5$ in a hydrodynamic model simulation with nuclear structure parameters close to those in Tab.~\ref{tab:1}~\cite{Xu:2021uar}. The impact of switching the event activity variable is generally smaller in more central collisions.  On the other hand, since  $\sigma_{\pT}/\lr{\pT}$ increases with centrality (or decreases with $\nchrec$), the difference in its ratio is negative, with a magnitude of $\sim0.01$ at $c= 0.2$, which increases to $\sim0.03$ at $c= 0.5$. The comparison with the AMPT simulations shows that the difference in the diffuseness parameter between the isobaric systems plays the dominant role. 

In the current study, $c$ and $\nchrec$ are analytically related in each collision system, and the two isobar ratios (at matching $c$ and at matching $\nchrec$) are different only because $c(\nchrec)_{\mathrm{Ru}}\neq c(\nchrec)_{\mathrm{Zr}}$ in Fig.~\ref{fig:0}. In general, event activity variables are not analytically related, such as the multiplicity determined at mid-rapidities and that determined at forward rapidities. As a concrete example, one can consider the relation between $\nchrec$ and $N_{\mathrm{EPD}}$, multiplicity in the STAR event-plane detector at $2<|\eta|<5$. In this case, one also has to take into account the fluctuations of $N_{\mathrm{EPD}}$ for events with fixed $\nchrec$, and vice versa. This effect, also known as volume fluctuations, may lead to additional differences in the isobar ratios~\cite{Zhou:2018fxx,Aaboud:2019sma} (besides that due to switching between $c$ and $\nchrec$). Specifically, events from $N_{\mathrm{EPD}}$ selection with an average of $\lr{\nchrec}=N_0$, may not have the same physics signal as events selected directly with a fixed $\nchrec=N_0$. The difference could be particularly large for multi-particle cumulant observables~\cite{Aaboud:2019sma,Sugiura:2019toh,Jia:2020tvb}. A study of this is in progress.

In summary, we have studied how the choice of event activity variable $x$ influences the observable ratio, $R_{\mathcal{O}} (x)$, of $^{96}$Ru+$^{96}$Ru to $^{96}$Zr+$^{96}$Zr collisions. Significant deviations of these ratios from unity have been attributed to the different collective nuclear structures  of $^{96}$Ru and $^{96}$Zr nuclei. In addition to directly influencing $\mathcal{O}$ in the two systems, the nuclear structure effects can also influence $x$  and hence cause additional differences in $R_{\mathcal{O}} (x)$. The latter contribution depends on the choice of event activity variable $x$ and the rate of change $\ln'(\mathcal{O}(x))$. Based on published STAR data, we calculate the difference $\Delta R_{\mathcal{O}}$ between the ratios calculated with two event activity variables: charged-particle multiplicity $\nchrec$ and centrality $c$.  The ratios of harmonic flow $v_n$ are nearly independent of the analysis methods when calculated at matching $\nchrec$ (i.e. $R_{v_n}(\nchrec)$) instead of at matching $c$ (i.e. $R_{v_n}(c)$), suggesting that nonflow effects are controlled by $\nchrec$ instead of centrality.  The isobar ratios for the CME sensitive observable ($\kappa_{112}$) are also closer to unity, when the event activity is quantified by $\nchrec$ instead of $c$, indicating that the former provides a better estimate of the background baseline for this observable. The AMPT model calculations are used to separate $\Delta R_{\mathcal{O}}$ into different effects of the nuclear structure parameters in the Woods-Saxon form. Surface diffuseness $a_0$ is found to have the largest influence, followed by nuclear radius $R_0$, and nuclear deformations only influence $\Delta R_{\mathcal{O}}$ in central collisions.  We also make a prediction on the expected $\Delta R$ for mean transverse momentum ($\lr{\pT}$) and its scaled-variance ($\sigma_{\pT}/\lr{\pT}$). The difference between the $\lr{\pT}$ ratios at matching $\nchrec$ and at matching centrality is sizeable compared with the total influence of the nuclear structure effects on $R_{\lr{\pT}}$. This difference needs to be taken into account explicitly in model calculations. 

{\bf Acknowledgements:}  We appreciate comments from Somadutta Bhatta. J. J. and C. Z. are supported by the U.S. Department of Energy under Grant No. DEFG0287ER40331. G. W. is supported by the U.S. Department of Energy under Grant No. DE-FG02-88ER40424.

\bibliography{deform}{}
\end{document}